\documentclass[submitting]{nst} 

\usepackage{subfigure,dcolumn}
\usepackage[T2A,T1]{fontenc}
\usepackage[russian,english]{babel}

\usepackage{listings}
\begin{document}

\title{Cluster Counting Algorithm for the CEPC Drift Chamber using LSTM and DGCNN}\thanks{This work was supported by National Natural Science Foundation of China (NSFC) (Nos. 12475200 and 12275296), Joint Fund of Research utilizing Large-Scale Scientific Facility of the NSFC and CAS (No. U2032114), Institute of High Energy Physics (Chinese Academy of Sciences) Innovative Project on Sciences and Technologies (Nos. E3545BU210 and E25456U210).}

\author{Zhe-Fei Tian}
\affiliation{Hubei Nuclear Solid Physics Key Laboratory, School of Physics and Technology, Wuhan University, Wuhan 430072, China}
\author{Guang Zhao}
\email[Corresponding author, ]{zhaog@ihep.ac.cn}
\affiliation{Institute of High Energy Physics, Chinese Academy of Sciences, Beijing 100049, China}
\author{Ling-Hui Wu}
\affiliation{Institute of High Energy Physics, Chinese Academy of Sciences, Beijing 100049, China}
\author{Zhen-Yu Zhang}
\email[Corresponding author, ]{zhenyuzhang@whu.edu.cn}
\affiliation{Hubei Nuclear Solid Physics Key Laboratory, School of Physics and Technology, Wuhan University, Wuhan 430072, China}
\author{Xiang Zhou}
\affiliation{Hubei Nuclear Solid Physics Key Laboratory, School of Physics and Technology, Wuhan University, Wuhan 430072, China}
\author{Shui-Ting Xin}
\affiliation{Institute of High Energy Physics, Chinese Academy of Sciences, Beijing 100049, China}
\author{Shuai-Yi Liu}
\affiliation{Institute of High Energy Physics, Chinese Academy of Sciences, Beijing 100049, China}
\author{Gang Li}
\affiliation{Institute of High Energy Physics, Chinese Academy of Sciences, Beijing 100049, China}
\author{Ming-Yi Dong}
\affiliation{Institute of High Energy Physics, Chinese Academy of Sciences, Beijing 100049, China}
\affiliation{University of Chinese Academy of Sciences, Beijing 100049, China}
\author{Sheng-Sen Sun}
\affiliation{Institute of High Energy Physics, Chinese Academy of Sciences, Beijing 100049, China}
\affiliation{University of Chinese Academy of Sciences, Beijing 100049, China}

\begin{abstract}
The particle identification (PID) of hadrons plays a crucial role in particle physics experiments, especially in flavor physics and jet tagging. The cluster-counting method, which measures the number of primary ionizations in gaseous detectors, is a promising breakthrough in PID. However, developing an effective reconstruction algorithm for cluster counting remains challenging. To address this challenge, we propose a cluster-counting algorithm based on long short-term memory and dynamic graph convolutional neural networks for the CEPC drift chamber. Experiments on Monte Carlo simulated samples demonstrate that our machine-learning-based algorithm surpasses traditional methods. It improves the $K/\pi$ separation of PID by 10\%, meeting the PID requirements of CEPC.
\end{abstract}


\maketitle

\section{Introduction} \label{sec:intro}
The Circular Electron Positron Collider (CEPC)~\cite{cepctdr, thecepcstudygroup2018cepc} is a large-scale collider facility proposed in 2012 after the discovery of the Higgs boson. It has a circumference of 100 km and two interaction points, which allows it to operate at multiple center-of-mass energies. 
Specifically, it serves as a Higgs factory at 240 GeV~\cite{an2019, yu2019, bai2020, tan2020}, facilitates $W^+W^-$ threshold scans at 160 GeV, and functions as a $Z$ factory at 91 GeV~\cite{shen2020, liang2019}. Furthermore, it can be upgraded to 360 GeV for a $t\bar{t}$ threshold scan. 
In the future, the CEPC can be upgraded to a proton-proton collider, enabling the direct exploration of new physics at a center-of-mass energy of approximately 100 TeV~\cite{gao2017, gao2021}. 
The primary scientific objective of the CEPC is to precisely measure the Higgs properties, particularly their coupling properties. Additionally, trillions of $Z \to q\bar{q}$ events produced by the CEPC offer an excellent opportunity to study flavor physics~\cite{zheng2020, li2022}.

The particle identification (PID) of hadrons is crucial in high-energy physics experiments, especially in flavor physics and jet tagging~\cite{zhu2023}. 
Particle identification can help suppress combinatorial backgrounds, distinguish between the final states of the same topology, and provide valuable additional information for jet flavor tagging. Future particle physics experiments, such as CEPC, require advanced detector techniques with PID performances that surpass current techniques.

\begin{figure*}[!htb]
    \centering
    \includegraphics[width=0.8\hsize]{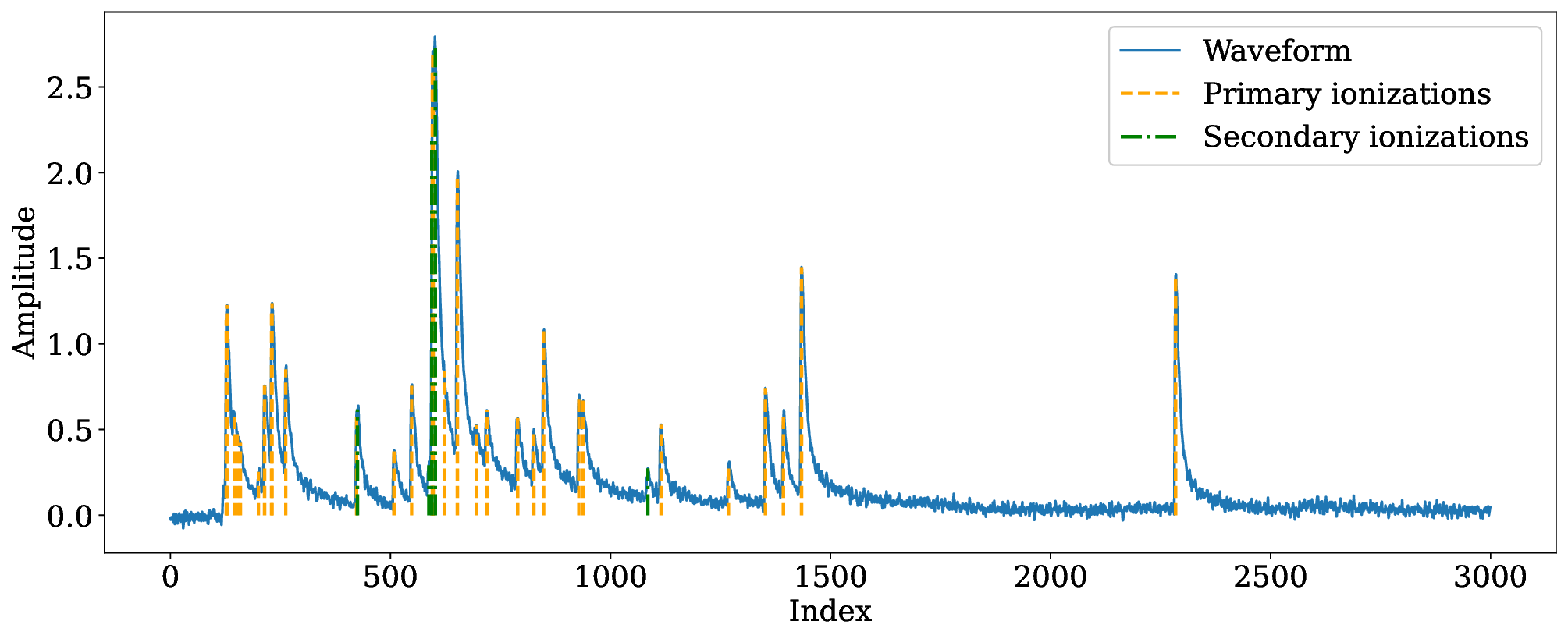}
    \caption{A waveform example of induced current on a sense wire of a drift chamber. The $x$-axis represents the index of the waveform, which is sampled over a time window of 2000 ns at a sampling rate of 1.5 GHz. Both primary and secondary ionizations contribute to the waveform. The orange lines indicate peaks from primary ionizations. The green lines indicate peaks from secondary ionizations. An effective reconstruction algorithm needs to efficiently and accurately count the number of primary ionizations in the waveform.}
    \label{fig:wf_example}
\end{figure*}

The drift chamber is a key detector in high-energy physical experiments. In addition to charged particle tracking, the drift chamber can also provide excellent PID while requiring almost no additional detector budget. In a drift chamber, PID is based on the ionization behavior of charged particles traversing the working gas. A well-established technique for identifying particles is the measurement of average ionization energy loss per unit length ($\mathrm{d} E /\mathrm{d} x$) of charged particles~\cite{CHARPAK1968262}. 
In a drift chamber cell, charged particles ionize the gas, creating a cascade of electrons that can be detected as primary signals. This type of ionization is called primary ionization and is a Poisson process. Moreover, some of these electrons can cause secondary ionization, leading to a Landau distribution $\mathrm{d} E /\mathrm{d} x$. The Landau distribution has an infinitely long tail and large fluctuations that limit the $\mathrm{d} E /\mathrm{d} x$-resolution~\cite{blum2008particle}. Figure \ref{fig:wf_example} shows an example signal waveform in a drift-chamber cell.

Alternatively, the cluster-counting technique directly measures the average number of primary ionizations per unit length in the waveforms processed by fast electronics, rather than $\mathrm{d} E /\mathrm{d} x$, which reduces the impact of secondary ionization ~\cite{IEEE4329616} and significantly improves PID performance. 
% The cluster-counting technique has the potential to double the resolution of PID. 
The cluster-counting technique has the potential to improve the resolution by a factor of two. 
Therefore, the cluster-counting technique, which is the most promising breakthrough in PID, has been proposed for future colliders for high-energy frontiers, such as the CEPC and the Future Circular Collider (FCC)~\cite{Abada2019}. A previous study on cluster counting for the BESIII upgrade demonstrated that the cluster-counting method exhibited superior PID performance compared with the $\mathrm{d} E /\mathrm{d} x$ method. This significantly enhanced PID performance for $\pi/K$, achieving a separation power that is approximately 1.7 times that of the $\mathrm{d} E /\mathrm{d} x$-method~\cite{Xin_2023}.

Reconstruction poses a significant challenge in cluster counting. 
An effective reconstruction algorithm must efficiently and accurately determine the number of primary ionizations in a waveform. However, the stochastic nature of ionization processes and the complexity of signals present substantial obstacles to developing reliable cluster-counting algorithms. 
In traditional methods, cluster-counting algorithms are typically divided into two stages: peak finding (detecting all peaks from both primary and secondary ionizations) and clustering (determining the number of primary ionizations among the detected peaks in the previous step). 
For derivative-based peak finding, the first and second derivatives of the waveform are computed, and signals are identified via threshold crossings. 
Unfortunately, derivative-based algorithms often fail to achieve state-of-the-art performance, especially in scenarios with high pile-up and noise levels.
In time-based clusterization, the average time differences between signals from different clusters tend to be larger than those within the same cluster. This information can be exploited to design peak-merging algorithms. However, due to the significant overlap in time difference distributions between inter-cluster and intra-cluster signals, these algorithms often suffer from low accuracy.
Machine learning (ML) is a rapidly advancing field in computer science that uses algorithms and statistical models to enable systems to improve their performance by learning from data. 
Neural networks, the most commonly used ML techniques, are computational models loosely inspired by the human brain and consist of interconnected layers. Recurrent neural networks (RNNs)~\cite{SHERSTINSKY2020132306} and graph neural networks (GNNs)~\cite{ZHOU202057} are particularly popular types of neural networks.
ML techniques have been widely applied in high-energy and nuclear physics.
For instance, in high-energy physics, the GNN-based ParticleNet algorithm was developed for jet tagging~\cite{Qu2020}, with applications to CEPC jet tagging~\cite{zhu2024}. In nuclear physics, ML techniques have been used to study phase transitions in nuclear matter governed by quantum chromodynamics (QCD)~\cite{ma2023,li2023}, and to analyze heavy-ion collisions across various energy scales\cite{he2023nst,he2023pma,gao2023}.
Machine learning has shown preliminary promise for handling large-scale data in high-energy physics. For cluster-counting algorithms, ML can leverage full waveform information and potentially uncover hidden features within the signal peaks. This problem can be modeled as a classification task, making it amenable to mature ML tools such as PyTorch~\cite{paszke2019pytorch} and PyTorch Geometric~\cite{Fey/Lenssen/2019}.

This paper presents an ML-based algorithm for cluster counting, optimized for a CEPC drift chamber. The remainder of the paper is organized as follows: 
Sec. \ref{sec:dataset} introduces the fast simulation method and the simulated samples used to train and test the ML-based algorithm. 
Sec. \ref{sec:method} details the ML-based cluster-counting algorithm. 
Sec. \ref{sec:performance} evaluates the 
performance of the ML-based algorithm and compares it with traditional methods. 
Sec. \ref{sec:conclusion} concludes the paper.

\section{Detector, Simulation and Data Sets} \label{sec:det-sim-sample}
\subsection{The CEPC Drift Chamber} \label{sec:detector}
In the design of the CEPC $4^{\rm th}$ conceptual detector, a drift chamber is proposed to be inserted between the silicon inner tracker (SIT) and the silicon external tracker (SET). 
This chamber primarily provides PID capability and enhances tracking and momentum measurements.

\begin{figure}[!htb]
    \centering
    \includegraphics[width=0.9\hsize]{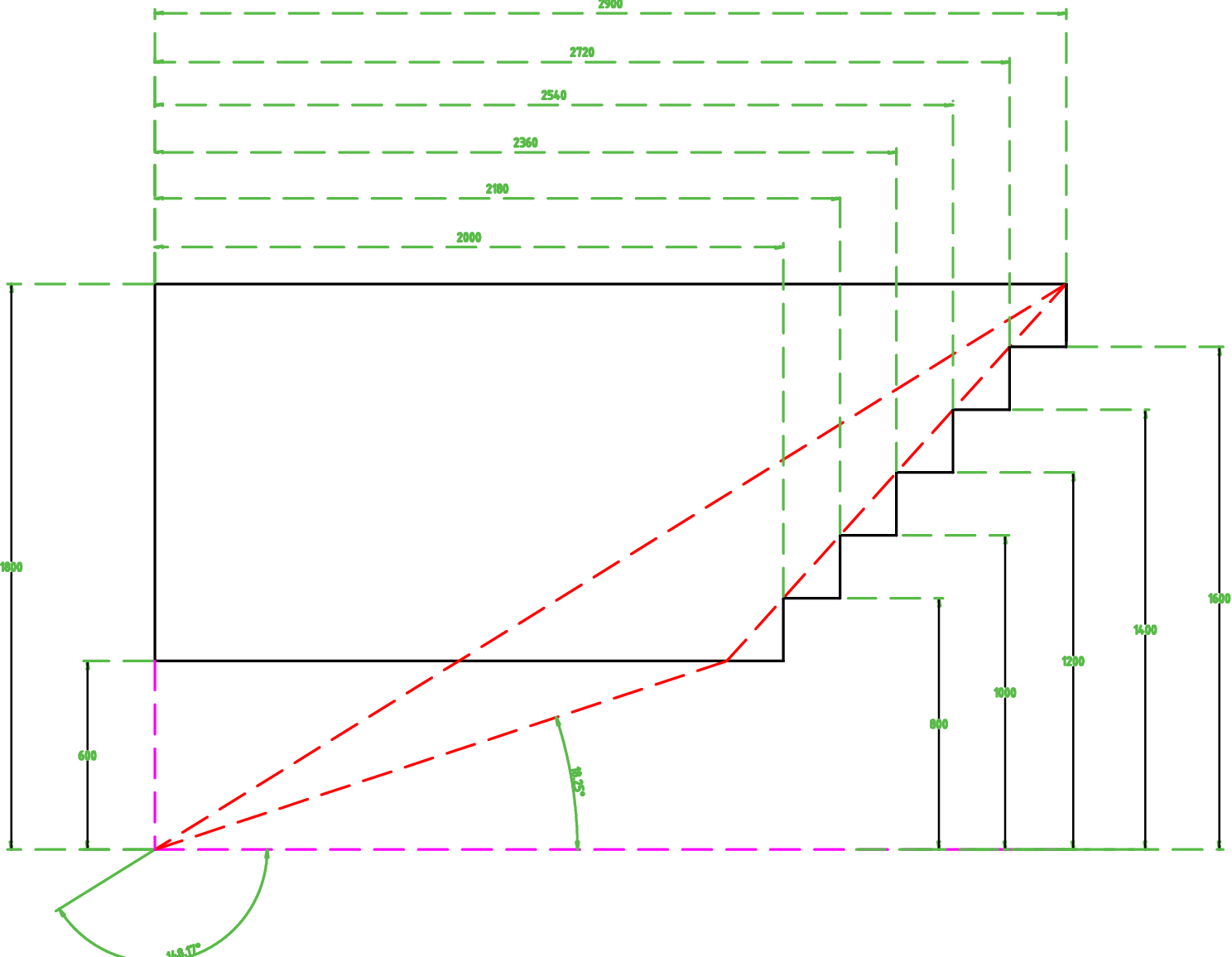}
    \caption{Schematic layout of one-fourth of the CEPC drift chamber. The black lines show the boundaries of the drift chamber.}
    \label{fig:dc}
\end{figure}

Based on the preliminary design, the chamber length was approximately 5800 mm, with a radial extent ranging from 600 to 1800 mm. 
The inner wall consisted of a carbon fiber cylinder, while the outer support featured a carbon fiber frame structure comprising eight longitudinal hollow beams and eight rings. 
These components were sealed with a gas envelope. 
The aluminum endplates were designed with a multistepped and tilted shape to minimize deformation caused by wire tension. 
A schematic of the drift chamber is shown in Fig. \ref{fig:dc}. 

The entire chamber comprises approximately 67 layers. To meet the requirements for PID capability and momentum measurements, a cell size of 18 mm $\times$ 18 mm was adopted. Each cell consists of a sense wire surrounded by eight field wires, forming a square configuration. The sense wires were 20 $\mu$m gold-plated tungsten wires, while the field wires were 80 $\mu$m gold-plated aluminum wires. To achieve a suitable primary ionization density, a gas mixture of 90$\%$ He and 10$\%$ iC$_{4}$H$_{10}$ was proposed.

\subsection{Simulation and Data Sets} \label{sec:dataset}
\begin{figure*}[!htb]
    \centering
    \includegraphics[width=0.9\hsize]{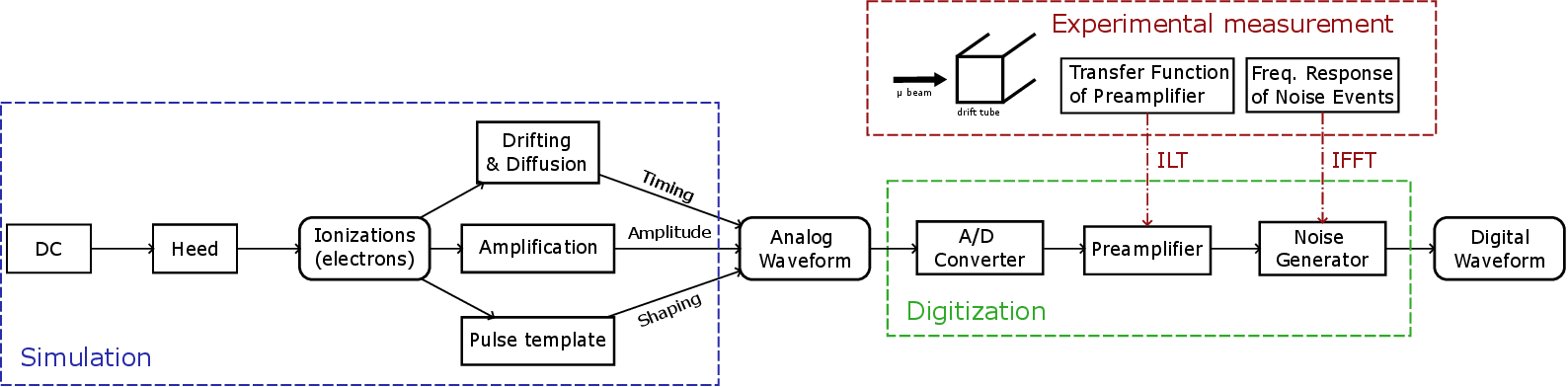}
    \caption{Simulation package for cluster-counting study. The package consists of simulation and digitization. The digitization takes input from the experimental measurement.}
    \label{fig:fast_sim}
\end{figure*}

A sophisticated first-principles simulation package was developed for cluster counting.
The package precisely simulates particle interactions and detector responses, creating realistic waveforms labeled with MC truth timing, which enables supervised training.
The simulation package consisted of two components: simulation and digitization. 
The geometry of the drift-chamber cells was constructed for the simulation. Ionizations of charged particles were generated using the Heed package. 
To reduce computational expense, the transportation, amplification, and signal creation processes for each electron were parameterized according to the Garfield++ simulation results, which output analog waveforms for drift chamber cells~\cite{PFEIFFER2019121}.
Data-driven electronic responses and noise were considered in digitization. The impulse response of the preamplifier was measured experimentally and convoluted with the waveform. 
Noise was extracted from experimental data using the fast Fourier transform and added to the signal via the inverse fast Fourier transform. 
The digitization outputs realistic digitized waveforms that exhibit good agreement with experimental data in terms of peak rise times and noise levels. 
A flowchart of the simulation is presented in Fig. \ref{fig:fast_sim}.

The simulation geometry is based on the design of the CEPC $4^{\rm th}$ conceptual detector. According to test beam experiments~\cite{CAPUTO2023167969}, the waveform exhibits a single-pulse rise time of approximately 4 ns, a noise level of 5$\%$, and a sampling rate of 1.5 GHz. 
Using the simulation package, MC samples with varying momenta were generated to train and test the neural network algorithm. Detailed information about the samples is presented in Tab. \ref{tab:datasets}.

\begin{table*}[!htb]
\centering
\caption{Summary of data sets used for training and testing ML-based cluster-counting algorithms.}
\label{tab:datasets}
\begin{tabular}{ccccc}
\toprule
{Purpose} & {Algorithm} & {Particle} & {Number of Events} & {Momentum (GeV/$c$)} \\
\midrule
Training & peak-finding & $\pi^\pm$ & $5\times10^5$ & $0.2-20.0$ \\
Testing & peak-finding & $\pi^\pm$ & $5\times10^5$ & $0.2-20.0$ \\
Training & Clusterization & $\pi^\pm$ & $5\times10^5$ & $0.2-20.0$ \\
Testing & Clusterization & $\pi^\pm$ & $1\times10^5 \times 7$ & $5.0/7.5/10.0/12.5/15.0/17.5/20.0$ \\
Testing & Clusterization & $K^\pm$ & $1\times10^5 \times 7$ & $5.0/7.5/10.0/12.5/15.0/17.5/20.0$ \\
\bottomrule
\end{tabular}
\end{table*}

\section{Methodology} \label{sec:method}
\subsection{Algorithm Overview}

An effective reconstruction algorithm for cluster counting must efficiently and accurately determine the number of primary ionizations in a waveform. As introduced in Sec. \ref{sec:intro}, the cluster-counting algorithm is typically decomposed into two steps: peak finding and clusterization. Peaks from both primary and secondary ionizations were detected, while clusterization discriminates primary ionizations from the peaks detected in the previous step.
The traditional peak-finding algorithm uses the first and second derivatives of a waveform~\cite{Zhao_2024}. Ionization electron pulses, characterized by a swift rise (mere nanoseconds) and prolonged decay (tens of nanoseconds), yield pronounced derivative values, facilitating peak identification. 
Higher-order derivatives enhance hidden peak detection but increase noise susceptibility. 
To mitigate high noise levels, preprocessing with low-pass filters, such as moving averages, is recommended before applying derivatives. 
For clusterization, a peak-merging algorithm was used. Electrons from a single primary cluster, which are typically spatially localized, exhibit proximate arrival times at the sensing wire, forming discernible clusters in the waveform. 
Timing information from peak detection aids in distinguishing primary and secondary electron signals. 
Nonetheless, due to potential overlap between electrons from distinct primary clusters, a precise peak-merging requirement is crucial for the clusterization algorithm.

The aforementioned traditional rule-based algorithms, which depend on incomplete raw hit information and human expertise, often fail to achieve state-of-the-art performance. 
In contrast, ML-based algorithms harness an abundance of labeled samples for supervised learning, directly extracting intricate data features. In the first step of cluster counting, a long short-term memory (LSTM) network is employed to discriminate between signals and noise. 
Both the primary and secondary ionization signals are detected during this step. 
The second step of the algorithm, clusterization, is achieved using a dynamic graph neural network (DGCNN). 
The DGCNN is used to classify whether a detected peak in the first step originates from primary ionization. 

\subsection{Peak-finding}
The peak-finding algorithm identifies all ionization peaks from a waveform.
To reduce complexity, waveforms are divided into sliding windows with a window size of 15 data points. 
For each sliding window, a label is added based on MC truth information. Labels can identify a signal candidate or a noise candidate, defining peak finding as a binary classification.

To process time-series data in sliding windows, an LSTM-based network is explored for the peak-finding algorithm. LSTM, a type of recurrent neural network (RNN), can process sequential data and has been successfully used in a range of applications~\cite{10.1162/neco.1997.9.8.1735}. 
RNNs are particularly effective for sequence modeling tasks, such as sequence prediction and labeling, because they utilize a dynamic contextual window that captures the entire history of the sequence. 
However, RNNs face limitations in processing long sequences effectively and are susceptible to issues related to vanishing and exploding gradients~\cite{rnn_difficulty, yuyong2019}.

The LSTMs have a unique architecture that includes memory blocks within a recurrent hidden layer. These memory blocks consist of memory cells and forget gates. 
Memory cells store the temporal state of the network through self-connections, while special multiplicative units, known as gates, regulate information flow. 
Each memory block includes an input gate to manage input activations in the memory cell, an output gate to control the output flow of cell activations, and a forget gate to scale the internal state of the cell before adding it as an input through self-recurrent connections, thereby adaptively forgetting or resetting the cell's memory~\cite{forget_gate, yuyong2019}.

The architecture of the LSTM-based peak-finding algorithm is summarized as follows:

\begin{itemize}
    \item An LSTM layer
    
    The LSTM layer is used for processing sequential data and capturing long-term dependencies between data points. This LSTM layer has one feature in the input data and 32 features in the hidden state.
    
    \item Two linear layers
    
    The neural network model consists of two linear layers that serve as fully connected layers. The first layer has an input size of 32 and an output size of 32. The second layer has an input size of 32 and an output size of 1. A sigmoid activation function~\cite{sigmoid} is applied to the output of the second layer to produce the final classification result.
\end{itemize}

\begin{figure}[!htb]
    \centering
    \includegraphics[width=0.9\hsize]{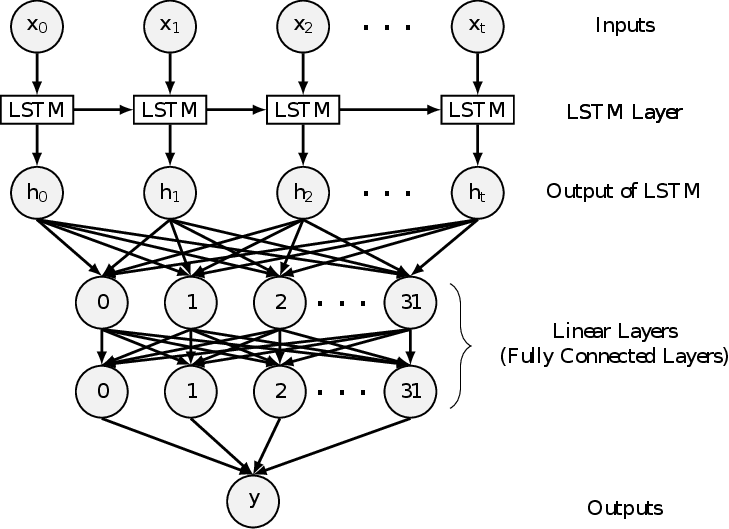}
    \caption{The neural network structure of the LSTM-based model for the peak-finding algorithm.}
    \label{fig:lstn_nn}
\end{figure}

Figure \ref{fig:lstn_nn} illustrates the network structure of the LSTM-based model used to train the peak-finding algorithm. 
The model was trained using a simulated sample of $\pi$ mesons, consisting of $5\times 10^5$ waveform events with momenta ranging from $\SI{0.2}{GeV}/c$ to $\SI{20}{GeV}/c$. 
After preprocessing, the data were divided into multiple batches, each with a batch size of 64, and the training process spanned 50 epochs.

Binary cross-entropy loss, a pivotal function for binary classification tasks, quantifies the discrepancy between true labels and predicted probabilities. This function effectively guides the model towards accurate predictions by handling cases where the output is a probability value between zero and one, making it particularly suited for our binary classification task. 
The Adam optimizer~\cite{adam} was adopted, with an initial learning rate of $10^{-4}$, which was reduced by a factor of 0.5 every 10 epochs. 
To further enhance algorithm performance, Optuna~\cite{Optuna}, a hyperparameter optimization framework, was employed to tune parameters such as the learning rate and network size.

\subsection{Clusterization}
After applying the LSTM-based peak-finding algorithm, all ionization signal peaks, including both primary and secondary peaks, were detected. 
A second algorithm, termed the clusterization algorithm, was then developed to determine the number of primary ionization peaks.

In principle, secondary ionization occurs locally with respect to primary electrons if the primary electrons possess sufficient energy. 
This proximity causes electrons from a single cluster to appear close together in the waveform, a property that can be exploited to design algorithms for distinguishing between primary and secondary electrons. 
As mentioned in Sec. \ref{sec:intro}, traditional algorithms for this purpose rely on combining adjacent peaks.

GNNs, which operate on graph-structured data, are well-suited for handling complex information. 
The key feature of GNNs is pairwise message passing, where graph nodes iteratively update their representations by exchanging information with their neighbors~\cite{gilmer2017}. For cluster counting, peak timing information is set as the node feature, while edges are initially connected based on timing similarities. GNNs can effectively learn the complex temporal structure of primary and secondary electrons through this message-passing mechanism.

\begin{figure*}[!htb]
    \centering
    \includegraphics[width=0.9\hsize]{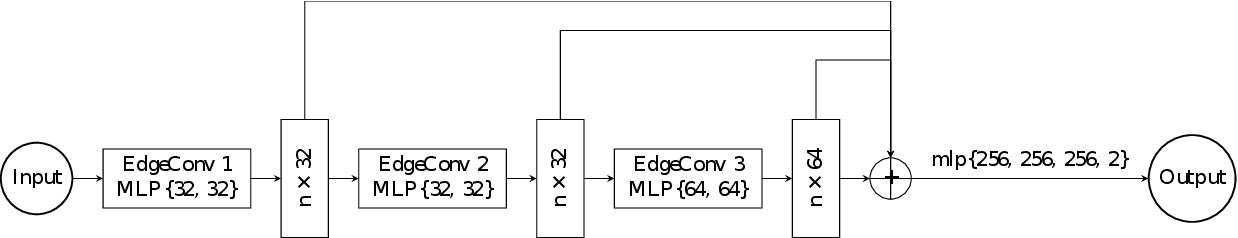}
    \caption{The neural network structure of the DGCNN-based algorithm for clusterization.}
    \label{fig:dgcnn_nn}
\end{figure*}

A DGCNN, a specialized type of GNN, was applied to the clusterization algorithm. 
DGCNNs are designed to learn from the local structure of point clouds, enabling high-level tasks such as classification and segmentation. 
The edge convolution layer, a critical component of DGCNNs, dynamically computes graphs at each network layer. 
This layer is differentiable and can be integrated into existing architectures. In this study, the timings of peaks detected during peak finding were represented as a graph. 
Each peak's timing was encoded as a node feature, while edge distances were defined by the temporal similarity between nodes.
Nodes were connected to their $k^{\rm th}$ nearest neighbors ($k$-NN)~\cite{dgcnn}. 
During training, nodes updated their features and connections through message passing, enabling the network to capture hidden local relationships between peaks and achieve better performance in classifying primary and secondary ionizations. 

The neural network architecture of the clusterization algorithm is summarized as follows:

\begin{itemize}
    \item Three dynamic edge convolution layers
    
    Three dynamic edge convolution layers process graph-structured data by dynamically creating edges between each node and its neighboring nodes, thereby capturing local information. A new graph is generated at each layer of the GNN based on the $k\text{-NN}$ approach~\cite{zhang2016}. The multi-layer perceptrons within the dynamic edge convolution layers map the number of input channels to the number of output channels.
    The features from three dynamic edge convolution layers were concatenated to get outputs with $32+32+64=128$ dimensions.
    
    \item A 4-layer multi-layer perceptron (MLP)
    
    Multi-layer Perceptron (MLP) is a type of feedforward neural network that consists of multiple layers of neurons connected in a sequential manner~\cite{Pinkus_1999}. This 4-layer MLP takes the concatenated output of the dynamic edge convolution layers as input. It has three hidden layers each with 256 neurons and 1 output layer with 2 channels. The dropout rate is set to 0.5, indicating that during training, each neuron in the network will have a 50\% probability of being randomly dropped in order to prevent overfitting and encourage the network to learn more robust features. Finally, the model applies a log-softmax activation function to the output of the MLP and returns the classification probabilities. 
\end{itemize}

Figure \ref{fig:dgcnn_nn} illustrates the neural network architecture for clusterization.
The model was trained using a pion sample containing $5\times10^5$ waveform events with momenta ranging from 0.2 GeV/$c$ to 20 GeV/$c$.
After preprocessing, the data were divided into multiple batches, each with a batch size of 128, and training was conducted over 100 epochs.

For this binary classification model, the negative log-likelihood loss function and the Adam optimizer were adopted, with an initial learning rate of $10^{-3}$, which was reduced by a factor of 0.5 every 10 epochs.
Hyperparameters, including the sizes of the three MLPs in the dynamic edge convolution layers and the MLP serving as a fully connected layer, were tuned using Optuna. The value of $k$ in $k$-NN, which determines how dynamic edge convolution layers establish relationships between nodes and their $k$ nearest neighbors, was optimized to 4.

\section{Performance} \label{sec:performance}

The two-step model was trained using supervised learning on a large number of waveform samples. To evaluate the model's generalization performance, it was applied to test samples.

\begin{figure}[!htb]
    \centering
    \includegraphics[width=0.9\hsize]{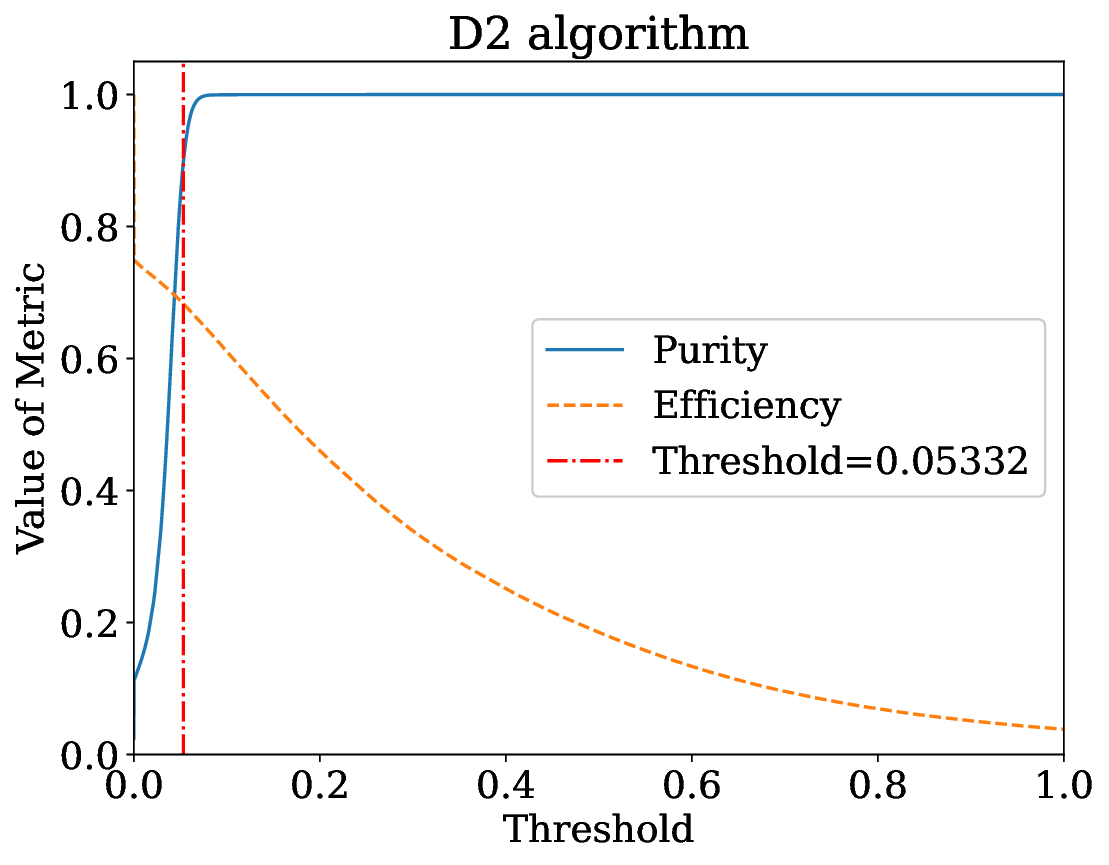}
    \includegraphics[width=0.9\hsize]{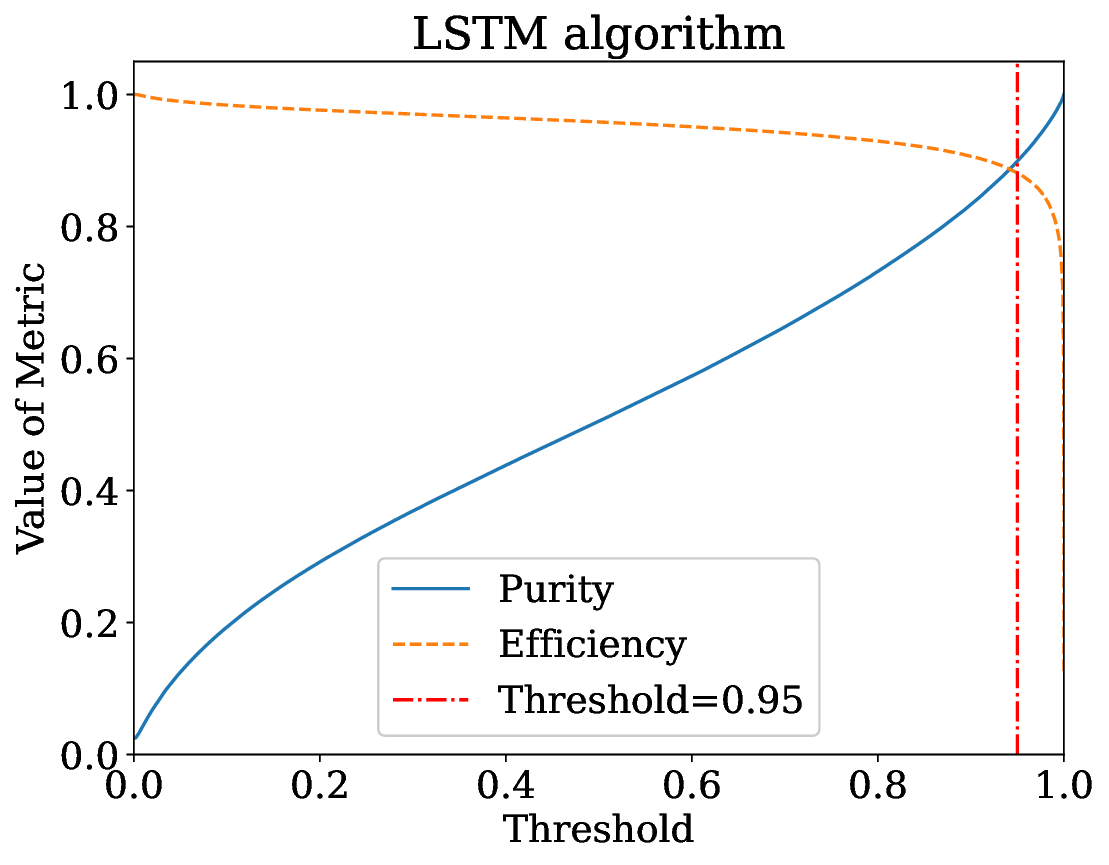}
    \caption{Purity and efficiency as a function of the threshold for derivative-based D2 and LSTM-based algorithm, respectively. The blue solid line is the purity curve, the orange dashed line is the efficiency curve, and the red dash dotted line is the optimized threshold. The threshold for the D2 algorithm acts on the second derivative. While the threshold for the LSTM algorithm applies to the predicted probability of the neural network, with a range of [0, 1]. Any candidate that surpasses this threshold, either from D2 or LSTM algorithm, is considered as an ionization peak.}
    \label{fig:pf_pr}
\end{figure}

\begin{figure}[!htb]
    \centering
    \includegraphics[width=0.9\hsize]{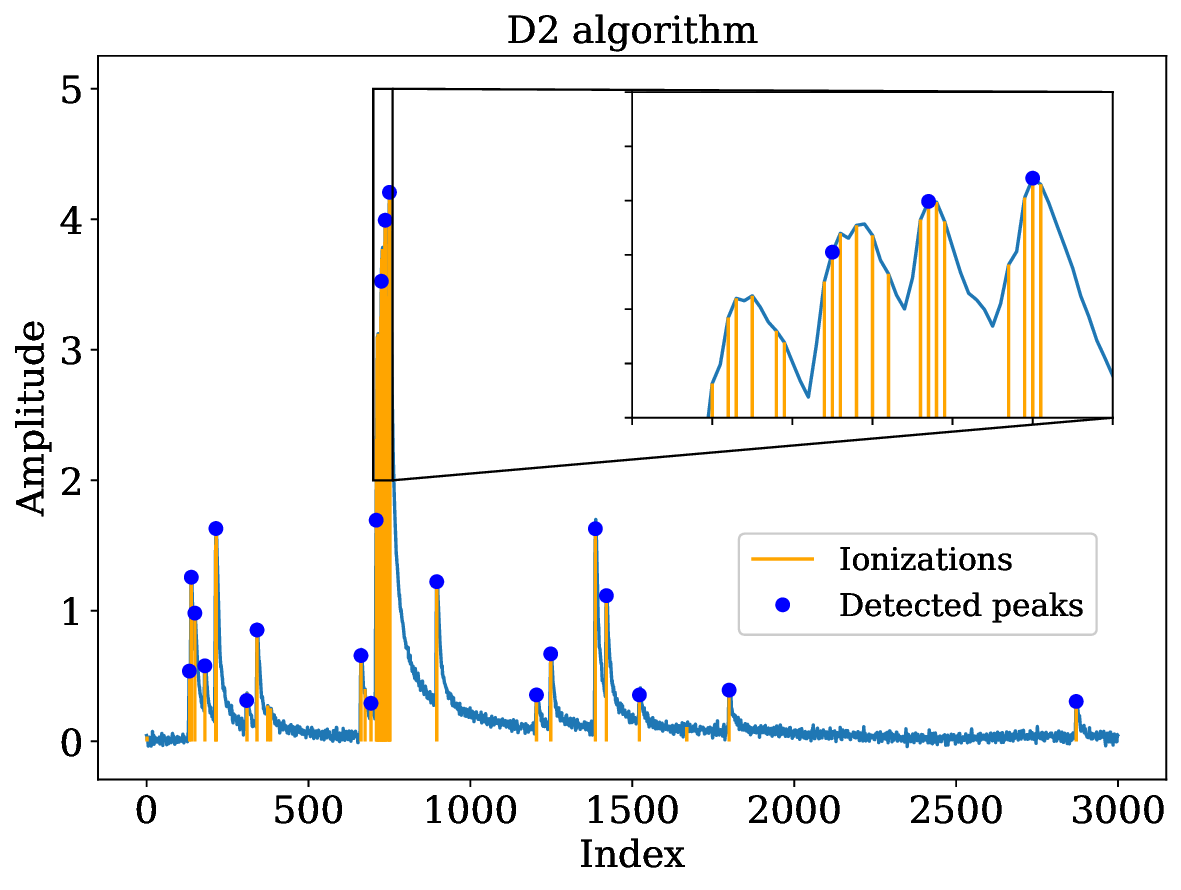}
    \includegraphics[width=0.9\hsize]{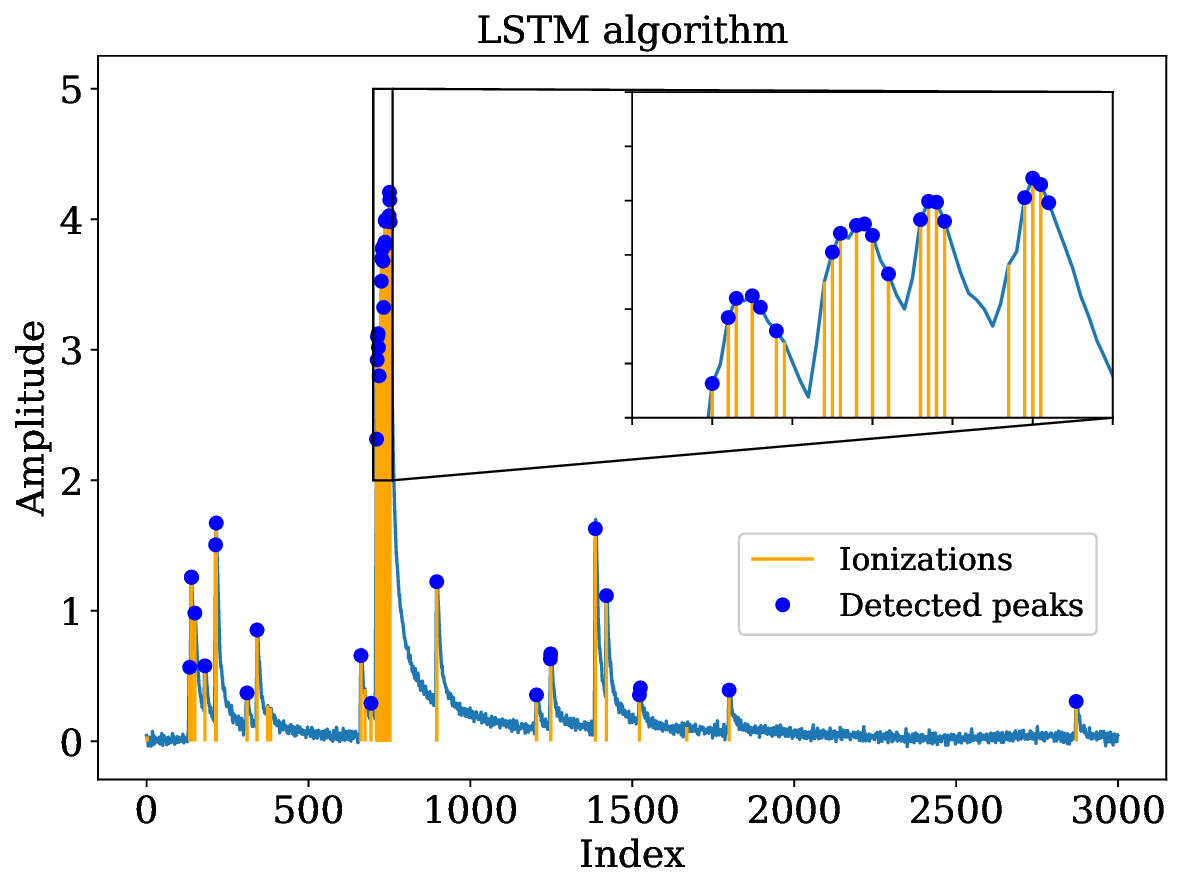}
    \caption{Applying the derivative-based D2 and LSTM-based peak-finding algorithms on a simulated waveform. The $x$-axis represents the index of the waveform, which is sampled over a time window of 2000 ns at a sampling rate of 1.5 GHz. The blue points are the detected peaks. The orange lines are the peaks from the MC truth. The zoomed figure shows that the LSTM-based algorithm detects the pile-up peaks more accurately and more efficiently than the D2 algorithm.}
    \label{fig:lstm_and_trad}
\end{figure}

For the peak-finding algorithm, both the LSTM-based algorithm and a traditional second-derivative-based (D2) algorithm served as classifiers. Their performance was evaluated using standard classifier metrics, including precision (purity) and recall (efficiency). Purity and efficiency are defined in terms of true positives (TP), false positives (FP), and false negatives (FN)~\cite{hossin2015review}, as shown in Eq. (\ref{eq:p_r}):
\begin{align}\label{eq:p_r}
\begin{split}
\text{Purity}&=\frac{\text{TP}}{\text{TP}+\text{FP}},\\
\text{Efficiency}&=\frac{\text{TP}}{\text{TP}+\text{FN}},
\end{split}
\end{align}
where TP is the number of correctly detected peaks, (TP+FP) is the total number of detected peaks, and (TP+FN) is the total number of peaks in MC truth of the waveform. The LSTM-based peak-finding algorithm was tested on a $\pi$ sample with momenta ranging from 0.2 GeV/$c$ to 20.0 GeV/$c$, consisting of $5\times10^5$ waveform events. 
Classifier purity and efficiency were evaluated by applying various probability thresholds. Figure \ref{fig:pf_pr} shows the purity and efficiency of the LSTM-based peak-finding algorithm and the traditional D2 algorithm as functions of the threshold. 
For the LSTM-based algorithm, a threshold of 0.95 yielded a purity of 0.8986 and an efficiency of 0.8820. For the D2 algorithm, the threshold was adjusted to match the LSTM-based algorithm's purity, yielding an efficiency of 0.6827 (Tab. \ref{tab:lstm_and_trad}).This demonstrates that the LSTM-based algorithm is significantly more efficient than the D2 algorithm, particularly in recovering pile-up events (Fig. \ref{fig:lstm_and_trad}).

\begin{table}[!htb]
\centering
\caption{The purity and efficiency comparison between LSTM-based algorithm and traditional D2 algorithm for peak-finding.}
\label{tab:lstm_and_trad}
\begin{tabular}{lcc}
\toprule
 & {Purity} & {Efficiency} \\
\midrule
LSTM algorithm & 0.8986 & 0.8820\\
D2 algorithm & 0.8986 & 0.6827 \\
\bottomrule
\end{tabular}
\end{table}

\begin{figure}[!htb]
    \centering
    \includegraphics[width=0.9\hsize]{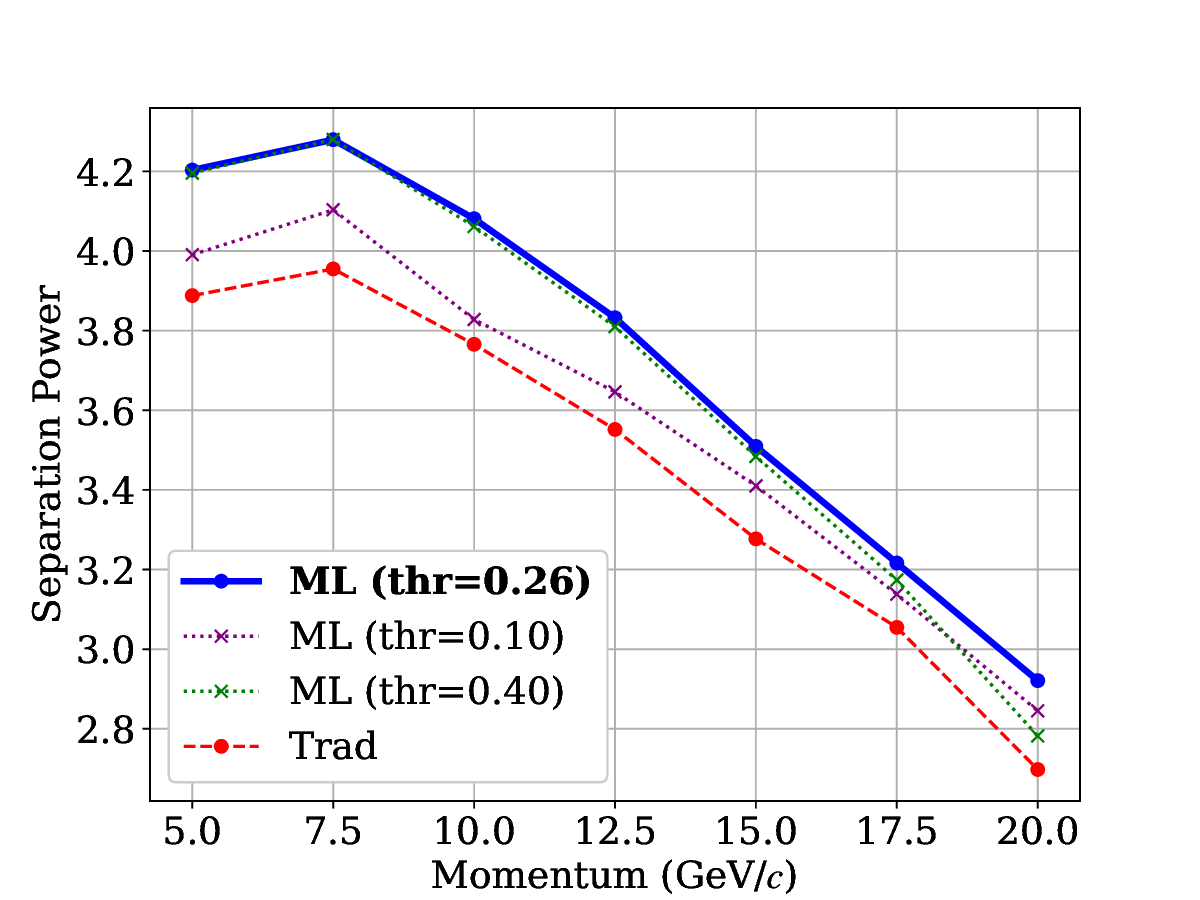}
    \caption{The $K/\pi$ separation power as a function of track momentum for a track length of 1 m. The red dashed line is from the traditional algorithm. The blue solid, violet dotted and green dotted lines are from the ML-based algorithm with thresholds of 0.26, 0.10, and 0.40. The blue solid line with a threshold of 0.26 achieves the overall best performance, which has a $K/\pi$ separation that is approximately 10\% better than that of the traditional algorithm.}
    \label{fig:sp}
\end{figure}

The clusterization algorithm was applied after peak finding to determine the number of primary clusters from the detected peaks. After implementing both the LSTM-based peak-finding and DGCNN-based clusterization algorithms, the number-of-cluster distribution for a charged particle was obtained, enabling the calculation of separation power for different types of charged particles. In this study, clusterization was achieved by performing node classification in the DGCNN.
To achieve optimal performance, the classifier threshold was tuned to maximize the $K/\pi$-separation power. The $K/\pi$-separation power is defined as 
\begin{equation}\label{eq:sp}
S=\frac{\left|\left(\frac{d N}{d x}\right)_\pi-\left(\frac{d N}{d x}\right)_K\right|}{\left(\sigma_\pi+\sigma_K\right) / 2},
\end{equation}
where $\mathrm{d} N /\mathrm{d} x_{\pi(K)}$ and $\sigma_{\pi(K)}$ represent the measured values and uncertainties in the number of primary ionizations per unit length for $\pi$ ($K$).
Optimization was performed using $K/\pi$ samples with fixed momenta of $p=5.0\text{ GeV}/c$, $7.5\text{ GeV}/c$, $10.0\text{ GeV}/c$, $12.5\text{ GeV}/c$, $15.0\text{ GeV}/c$, $17.5\text{ GeV}/c$, and $20.0\text{ GeV}/c$. The solid blue, dashed violet, and dashed cyan lines in Fig. \ref{fig:sp} illustrate the $K/\pi$ separation power for different thresholds. According to the optimization, the model achieves the best overall performance with a threshold of 0.26.

\begin{figure}[!htb]
    \centering
    \includegraphics[width=0.9\hsize]{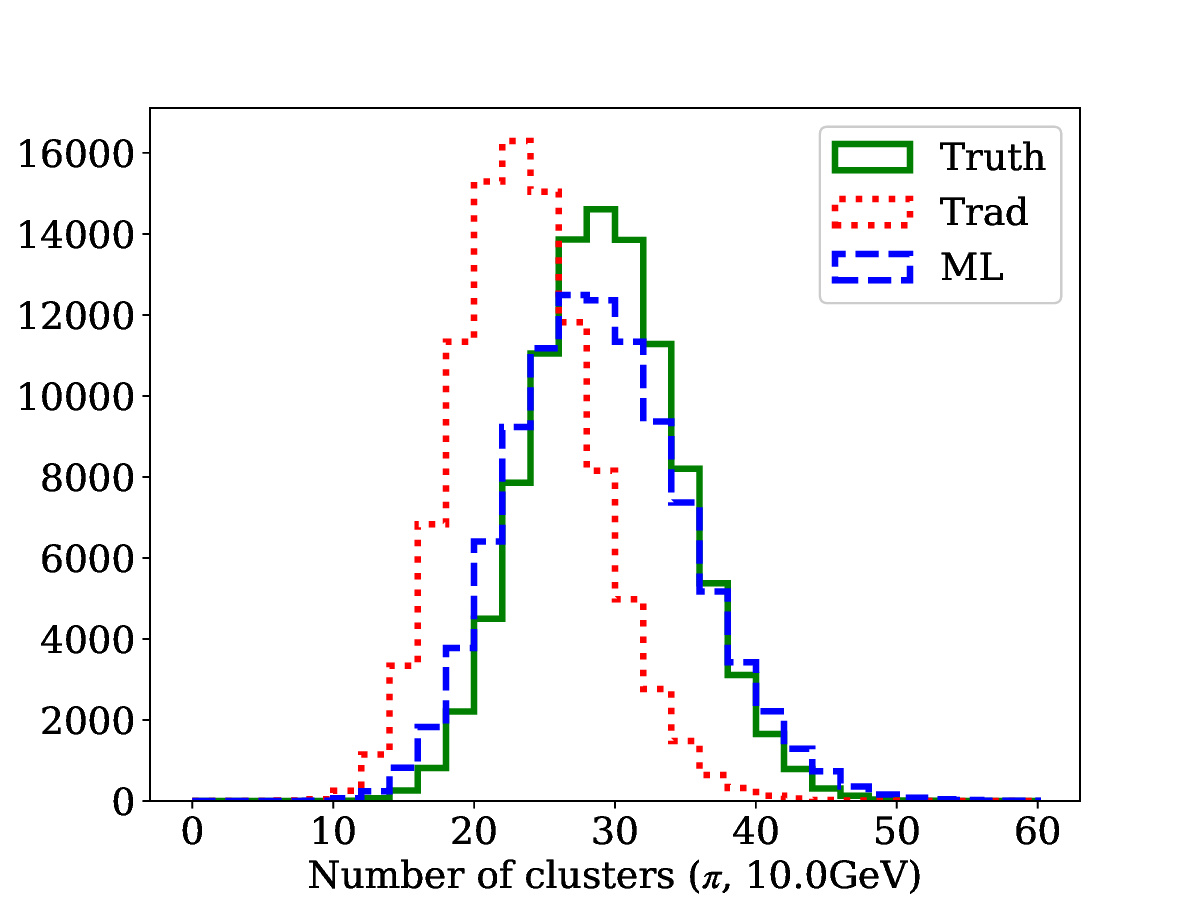}
    \caption{The number-of-cluster distribution from MC truth (solid green), reconstruction by a traditional algorithm (dotted red) and reconstruction by an ML-based algorithm (dashed blue) for a $10\text{ GeV}/c$ pion sample.}
    \label{fig:ncls_10GeV}
\end{figure}

\begin{figure}[!htb]
    \centering
    \includegraphics[width=0.9\hsize]{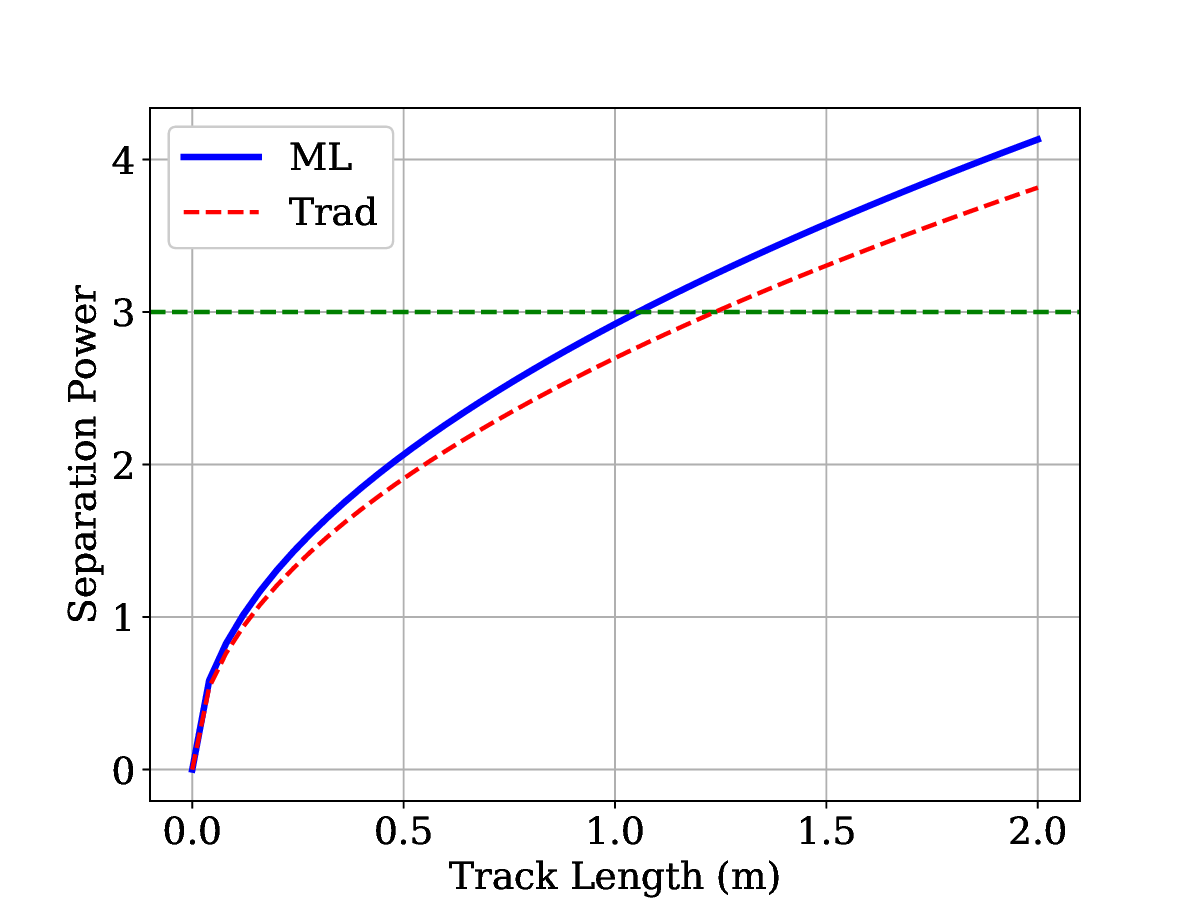}
    \caption{The $K/\pi$ separation power as a function of track length ($L$) at $20\text{ GeV}/c$. The curve is extrapolated from the value at $L=1{\rm m}$ by $\sqrt{L}$. The red dashed line is from the traditional algorithm. The blue solid line is from our ML-based algorithm with two steps. The green dashed line shows the target of $3\sigma$ separation power.}
    \label{fig:length}
\end{figure}

Using the optimized threshold, Fig. \ref{fig:ncls_10GeV} compares the number-of-cluster distributions derived from the MC truth, the traditional algorithm, and the DGCNN-based algorithm. 
The mean value of the number-of-cluster distribution obtained from the ML-based algorithm closely aligns with the MC truth, demonstrating that the ML-based algorithm achieves higher efficiency than traditional approaches. 
Figure \ref{fig:sp} presents the $K/\pi$ separation powers at various momenta for a track length of 1 m using different algorithms.
The ML-based cluster-counting algorithm shows approximately a 10\% improvement in separation power across all momenta compared to traditional methods.
Since separation power scales with the square root of the track length, this performance improvement corresponds to an effective increase of about 20\% in the detector radius when using the traditional algorithm. 
This enhancement could significantly reduce the overall cost of the detector. Detailed numerical results are listed in Tab. \ref{tab:particle_momentum}. Additionally, $K/\pi$ separation power for a track with $p=20\text{ GeV}/c$ was extrapolated to different track lengths starting at 1 m. Figure \ref{fig:length} shows the $K/\pi$ separation power as a function of track length.
The CEPC design requires a 3$\sigma$ $K/\pi$ separation for momenta up to 20 GeV/$c$. Using the ML-based reconstruction algorithm, the current drift chamber design, with a radius ranging from 600 mm to 1800 mm, meets the necessary PID requirements.

\begin{table*}[htb!]
\centering
\caption{Efficiency and separation power for charged $K$ and $\pi$ at various momenta and for different algorithms. The threshold of the ML-based algorithm is optimized as 0.95 for the LSTM-based peak-finding algorithm and 0.26 for the DGCNN-based clusterization algorithm. 
The efficiency is defined as the ratio of the number of reconstructed clusters to the number of MC truth clusters.}
\label{tab:particle_momentum}
\begin{tabular}{llccccccc}
\toprule
& & \multicolumn{7}{c}{{Momentum({GeV}$/$\textit{c})}} \\
\cmidrule(r){3-9}
Algorithm & Metric & {5.0} & {7.5} & {10.0} & {12.5} & {15.0} & {17.5} & {20.0}\\ 
\midrule
ML-based algorithm & $\pi^\pm$ efficiency & 1.003 & 1.001 & 0.999 & 0.999 & 0.998 & 0.998 & 0.999\\
 & $K^\pm$ efficiency & 1.014 & 1.011 & 1.010 & 1.008 & 1.006 & 1.004 & 1.003\\
 & $K/\pi$ separation power & 4.203 & 4.279 & 4.081 & 3.832 & 3.509 & 3.216 & 2.921 \\
\midrule
Traditional algorithm & $\pi^\pm$ efficiency & 0.814 & 0.808 & 0.803 & 0.801 & 0.801 & 0.800 & 0.800\\
 & $K^\pm$ efficiency & 0.837 & 0.830 & 0.824 & 0.820 & 0.817 & 0.814 & 0.812\\
 & $K/\pi$ separation power & 3.888 & 3.954 & 3.765 & 3.550 & 3.277 & 3.054 & 2.697 \\
\bottomrule
\end{tabular}
\end{table*}

\section{Conclusion} \label{sec:conclusion}

In this study, we developed a cluster-counting algorithm that incorporates both peak-finding and clusterization algorithms based on ML. Our approach offers several advantages over traditional cluster-counting methods. In particular, our peak-finding algorithm demonstrated better efficiency than the derivative-based algorithm. The clusterization algorithm provides a Gaussian-distributed number of clusters and achieves an efficiency close to that of the ground truth (MC truth). The entire cluster-counting algorithm outperformed traditional methods, showing a 10 \% improvement in the $K/\pi$ separation power. This level of PID performance with ML-based algorithms is approximately equivalent to having a 20\% larger detector size than traditional algorithms. With such performance, the current design of the CEPC drift chamber meets the necessary PID requirements. Furthermore, the critical role of ML-based algorithms in cluster counting suggests their potential applications in future high-energy physics experiments.

\noindent \textbf{Data availability} The data that support the findings of this study are openly available in Science Data Bank at \href{https://doi.org/10.57760/sciencedb.16322}{https://doi.org/10.57760/sciencedb.16322} and \href{https://cstr.cn/31253.11.sciencedb.16322}{https://cstr.cn/31253.11.sciencedb.16322}.

\end{document}